# Local atomic configuration control of superconductivity in the undoped pnictide parent compound BaFe$_2$As$_2$

Jong-Hoon Kang[1], Philip J. Ryan[2,5], Jong-Woo Kim[2], Jonathon Schad[1], Jacob P. Podkaminer[1], Neil Campbell[3], Joseph Suttle[3], Tae Heon Kim[1], Liang Luo[4], Di Cheng[4], Yesusa G. Collantes[6], Eric E. Hellstrom[6], Jigang Wang[4], Robert McDermott[3], Mark S. Rzchowski[3], Chang-Beom Eom[1]*

[1]Department of Materials Science and Engineering, University of Wisconsin-Madison, Madison, Wisconsin 53706, USA

[2]Advanced Photon Source, Argonne National Laboratory, Argonne, Illinois 60439, USA

[3]Department of Physics, University of Wisconsin-Madison, Madison, Wisconsin 53706, USA

[4]Department of Physics and Astronomy, Ames Laboratory, Iowa State University, Ames, Iowa 50011, USA

[5]School of Physical Sciences, Dublin City University, Dublin 9, Ireland

[6]Applied Superconductivity Center, National High Magnetic Field Laboratory, Florida State University, 2031 East Paul Dirac Drive, Tallahassee, FL 32310, USA

**Emergent superconductivity is strongly correlated with the symmetry of local atomic configuration in the parent compounds of iron-based superconductors**. **While chemical doping or hydrostatic pressure can change the local geometry, these conventional approaches do not provide a clear pathway in tuning the detailed atomic arrangement predictably, due to the parent compounds complicated structural deformation in the presence of the tetragonal-to-orthorhombic phase transition. Here, we demonstrate a systematic approach to manipulate the local structural configurations in BaFe$_2$As$_2$ epitaxial thin films by controlling two independent structural factors orthorhombicity (in-plane anisotropy) and tetragonality (out-of-plane/in-plane balance) from lattice parameters. We tune superconductivity without chemical doping utilizing both structural factors separately, controlling local tetrahedral coordination in designed thin film heterostructures with substrate clamping and bi-axial strain. We further show that this allows quantitative control of both the structural phase transition, associated magnetism, and superconductivity in the parent material BaFe$_2$As$_2$. This approach will advance the development of tunable thin film superconductors in reduced dimension.**

*eom@engr.wisc.edu

A pervasive issue affecting the fundamental understanding and practical application of high temperature superconductors is identifying the factors that determine the maximum critical temperature ($T_{c,max}$) for a given material family. In the case of Fe-based superconductors, attention has primarily focused on the inter-relation between superconductivity and other broken-symmetry phases found in the phase diagram, principally magnetic order[1-8]. Even though the physical origin of the symmetry-breaking ground state remains controversial, structural factors, possibly associated with the above relationship, have been shown to play an important role in determining $T_{c,max}$ and broken symmetries[9-17]. Several early investigations of a variety of optimally doped, tetragonal, bulk Fe-pnictide superconductors revealed an apparent correlation between $T_{c,max}$ and the proximity to perfect tetrahedral coordination of the Fe ions, which can control effective hopping parameters for electrons moving in the associated Fe-As planes[18-20]. For underdoped compositions, however, the presence of a tetragonal-to-orthorhombic phase transition makes it considerably more difficult to assess the impact of these structural effects on $T_c$.

Structural distortions and superconductivity have been induced in the undoped pnictide parent compound by either chemical substitution or application of large hydrostatic pressure[9,21-25]. It has been suggested that emergent superconductivity may be associated with local or global strains that change the local atomic structure, affecting the tetrahedral bond angles $\alpha$, $\beta$ and $\gamma$. However, these external stimuli do not provide an ideal approach to assess the relative roles played by structural distortions and associated structure-property relations because the internal distortion of the tetrahedral geometry is not predictable. A desired approach is to determine the orthogonal factors to suppress the orthorhombic distortion and change the tetragonal unit cell independently, and hence control the detailed bond angles of sub-unit cell structures in the parent material.

Here we use epitaxial strain in designed thin film heterostructures to control crystal symmetry such as in-plane (orthorhombicity) and in-plane/out-of-plane anisotropies (tetragonality), and by doing so the sub-lattice structure can be precisely manipulated. For this, the nominally non-superconducting parent compound $BaFe_2As_2$ (Ba-122) is utilized to grow epitaxial thin films on different substrates[26] with different film thicknesses[15], and correlate changes in $T_c$ with structural factors. Structural control of Ba-122 thin films was through the epitaxial relationship of the film with the substrate (see Methods and Supplementary Figure S1). As indicated schematically in Fig. 1, we employ two strategies to control the degree of tetragonality $c/a$ above the structural phase transition (which occurs at a temperature $T_s$), and the orthorhombicity $\delta = (a-b)/(a+b)$ for temperatures below $T_s$. First, different cubic substrates ($SrTiO_3$, $CaF_2$ and LiF) and growth conditions directly tune $c/a$ above $T_s$ without breaking any symmetry. Second, by growing films with different thicknesses, the degree of orthorhombicity $\delta$ below $T_s$ is effectively tuned. In particular, for very thin films, $\delta$ is suppressed due to a "clamping" effect arising from epitaxy with the cubic substrate. This relaxes as the film thickness is increased (approaching a free-standing bulk crystal), leading to larger values of $\delta$ for thicker films. Hence substrate type and film thickness provide controls of the As-Fe-As bond angles $\alpha$, $\beta$ and $\gamma$ even in the orthorhombic phase.

## Results

### *Influence of orthorhombicity and tetragonality*

Tetragonality and orthorhombicity provide a straightforward way to manipulate bond angles (Figure 1). In the tetragonal structure above $T_s$, there are only two unique bond angles $\alpha$ and $\beta = \gamma$ (see Methods). In the orthorhombic state below $T_s$, $\beta$ differs from $\gamma$ as a result of broken rotational symmetry arising from anisotropy of the in-plane lattice constants. The bond angle $\alpha$ is associated with the next-nearest-neighbor interaction of Fe $3d$ orbitals, while $\beta$ and $\gamma$ are linked with nearest-neighbor hopping, determining antiferromagnetic ordering[19,27] (see Supplementary Figure S2). We control these bond angles, and by these the electronic and magnetic properties, with film tetragonality and orthorhombicity.

### *Control of orthorhombic distortion by the clamping effect*

Figure 2 shows that superconductivity can be enhanced by suppression of orthorhombicity. High-resolution synchrotron XRD demonstrates that a single sharp peak is observed above the structural transition temperature $T_s$ in Fig. 2a. Regardless of film thickness, growing films on LiF substrates results in identical sharp peak positions for temperatures just above $T_s$ (~160 K), which implies constant tetragonal lattice parameters of the films. Upon cooling below $T_s$, the (228) reflection is broadened with a peak intensity drop due to peak splitting associated with the orthorhombic transition as observed in bulk[28,29]. The splitting is temperature-dependent, increasing with decreasing temperature until reaching a low-temperature plateau. Its overall magnitude is reduced in thinner films due to substrate clamping. We fit to the XRD data to determine the structural factor of orthorhombicity, $\delta = (a-b)/(a+b)$, and use its low-temperature plateau value to quantitatively compare between films, and correlate with electronic properties (see Methods). Figure 2b shows the temperature-dependent $\delta$, with a transition to the low-temperature plateau that is broader in these thin films than in bulk crystals[6]. The overall reduction of $\delta$ with decreasing film thickness is clearly shown. The low-temperature saturated $\delta$ value is reduced and the onset of orthorhombicity is shifted to lower temperature as film thickness is reduced[30].

As shown in Fig. 2b and 2c, there is a resistive anomaly near the orthorhombic structural transition. We assign $T_s$ to the onset of the upturn in the first derivative of the resistivity with respect to temperature (d$\rho$/dT), and use this value to characterize the orthorhombic transition (see Supplementary Figure S3). This provides a more precise comparison temperature than that obtained from the XRD data. We confirmed that the onset of orthorhombic $\delta$ is consistent with this resistive $T_s$ (Fig. 2b), as in bulk materials[29-31]. The structural transition measured by the orthorhombicity shows a broad temperature dependence, unlike the near 1$^{st}$ order transition observed in Ba-122 single crystals[29-31]. We probe the magnetic phase transition temperature ($T_n$) with reflection optical polarimetry[32] (see Methods). Here the measured ellipticity signal, which arises from the two-fold in-plane anisotropy of the refractive index, sets in below $T_s$, consistent with the emergence of orthorhombicity, and exhibits a kink at $T_n$ due to the magnetic feedback effect[32]. $T_n$ does not coincide with $T_s$ in our undoped films as shown in anisotropic characteristics revealed by ellipticity and structural orthorhombicity measurements[33] (see Supplementary Figure S4c), similar to the behavior of the electron doped parent compound[34,35].

*Control of the tetragonal structure with bi-axial strain*

Strain-controlled tetragonality is another structural factor controlling bond angles. As the $c/a$ ratio is increased by compressive strain, bond angles $\alpha$ and $\beta$ change linearly and approach a single value (Figure 3c). Note that we need to consider bond angle $\gamma$ as well because orthorhombicity is not completely suppressed. We control the $c/a$ by film strain, and As-Fe-As bond angles are systematically controlled by $c/a$ ratio, approaching the optimal angle of 109.5°.

Since the clamping effect differs on each substrate, the same orthorhombicity for films of differing tetragonality can be obtained by varying the film thickness. This allows separation of effects due to tetragonality and orthorhombicity. We confirmed a set of films on different substrates with identical orthorhombicity but varying tetragonality through measurements of temperature-dependent $\delta$ (Figure 3a and 3b). The out-of-plane lattice constant varies in this set, determined by the in-plane lattice constants through elasticity. Thus substrate clamping provides tetragonality control through the substrate in-plane lattice constant. For instance, x-ray diffraction confirms that Ba-122 films grown on dissimilar substrates have different in-plane lattice constants and also different tetragonality ($c/a$) due to the film strain. In particular, fluoride substrates provide more compressive strain than oxides, due to higher thermal contraction and resulting thermal strain.

*Influence of controlled structural distortions on $T_s$ and $T_c$*

We have demonstrated that tetragonality and orthorhombicity can be controlled in the undoped $BaFe_2As_2$. Here we show that these structural distortions determine the orthorhombic structural phase transition temperature $T_s$ and the superconducting phase transition temperature $T_c$, inducing superconductivity in this undoped parent compound. These results are summarized in Fig. 2d and 3e. Normalized resistivity of individual samples shows a distinct superconducting transition, and the onset of $T_c$ is obtained from the resistivity curve (Figure 3d). Note that, as discussed above, the 11 nm thick Ba-122 film grown on LiF at 680°C has zero resistance at 0.4K (inset of Figure 2c).

Figure 3e shows that superconductivity with an onset temperature as high as 27K can be induced by a suppression of orthorhombicity by substrate clamping at low thickness. Figure 2d shows that films of fixed orthorhombicity but different tetragonality have the same $T_s$, but dramatically varying $T_c$. We can summarize our results in terms of orthorhombicity, tetragonality, the orthorhombic structural phase transition temperature $T_s$ and the superconducting phase transition temperature $T_c$. In the range we investigated, tetragonality enhances superconductivity and orthorhombicity degrades superconductivity. Tetragonality does not strongly influence $T_s$, while orthorhombicity correlates with an increase in $T_s$. Although structural deformations and their interaction with each other, with electronic and magnetic properties, and with spatial gradients can be quite complex, our results point to clear correlations. We have shown that the reduction of in-plane lattice anisotropy ($a$ and $b$) increases the superconducting transition temperature.

The constant $T_s$ in films with varying tetragonality but identical orthorhombicity is reflective of the nematic transition, driven by either a spin or orbital mechanism. If a spin or orbital mechanism drives the nematic transition, its essential coupling to the orthorhombic structural distortion would connect a larger low-temperature orthorhombicity with a stronger nematicity and associated higher transition temperature. We argue that the dependence of superconducting $T_c$ on both orthorhombicity and tetragonality is related to the Fe-As bond angles. We found $T_c$ is enhanced by proximity to all bond angles approaching 109.5° (inset of Figure 3e). Despite lacking experimental realization of perfect 109.5° tetrahedral angles, we have shown that reduction of asymmetry between $\alpha$ and $\beta$ (or $\gamma$) enhances the superconducting $T_c$.

Superconductivity emerges as reduced film thickness decreases orthorhombicity, which to our knowledge has not been previously reported on insulating substrates. The enhancement of $T_c$ is tuned by film thickness due to the substrate clamping effect (inset of Figure 2d). The film with the smallest orthorhombicity becomes fully superconducting with zero resistance, as shown in the inset of Figure 2c (see Supplementary Figure S5). $T_s$ and onset $T_c$ (due to the broad transition) were plotted as functions of $\delta$ to show the correlation of orthorhombicity factor ($\delta$) with superconducting $T_c$ (Figure 2d). It is intriguing to see that reduction of $\delta$ is correlated to lower $T_s$ and higher $T_c$.

## *Visualizing the parent Ba-122 phase diagram*

Although our films do not cover the complete range of single-phase structural distortions, we visualize a phase diagram of parent Ba-122, guiding how to manipulate the As-Fe-As configuration by controlling two separate structural variables (Figure 4). With designed heterostructures, the selection of substrate templates and film thickness enables individual control parameters to achieve desired structures. The phase diagram summarizes that decoupled structural factors describing tetrahedral geometry can be systematically controlled so that superconductivity emerges in parent Ba-122. Superconductivity is enhanced by reducing broken symmetry phases as orthorhombicity decreases due to clamping (along red line) and by approaching a regular tetrahedral coordination as compressive strain tunes tetragonality (along blue line).

## *Conclusions*

We experimentally demonstrate a systematic approach to precisely control tetrahedral atomic configurations in parent compounds of Fe-based superconductors through control of independent structural factors orthorhombicity and tetragonality. Our observations are quantitatively explained using atomic-scale detailed structures and concomitant superconducting properties in the thin film heterostructures. This confirms our structural approach is applicable to induce and enhance the superconductivity in ultrathin films without chemical doping in a predictable way. A pure monolayer FeSe with our designed heterostructures would be a good platform to further probe structure-property relationships, and to explore unconventional superconductivity in low-dimensional heterostructure films[36,37]. Further, our ultrathin $BaFe_2As_2$ films could break new ground in non-equilibrium materials discovery using terahertz electromagnetic radiation in the superconducting[38,39] and nematic states[40]. We believe our study will provide a path toward tunable low dimensional superconductivity by atomic structure design.

## Methods

**Epitaxial growth.** Parent compound Ba-122 thin films were grown on various (001) oriented single-crystal substrates by pulsed laser deposition with a KrF (248 nm) ultraviolet excimer laser at 650-740°C. High-quality epitaxial films were synthesized on strontium titanate (STO), calcium fluoride ($CaF_2$), and lithium fluoride (LiF), as shown in Supplementary Figure S1. Temperature-dependent in-plane lattice change of the films is driven by the thermal contraction coefficient of substrate and different strain states obtained by temperature-dependent lattice mismatch (see Supplementary Figure S7 and Figure S8). We find fluoride substrates with higher thermal expansion coefficient (CTE) are more effective in providing compressive strain compared to oxide substrates. Note that the CTEs of LiF, $CaF_2$, and STO are $3.5 \times 10^{-5}$/K, $1.9 \times 10^{-5}$/K, $9.4 \times 10^{-6}$/K at 300K, respectively, which are all higher than that of Ba-122 with $1.0 \times 10^{-6}$/K. Due to the low melting point of LiF, growth temperature is below 700°C, but two different strain states are successfully provided by higher CTE grown at 650°C and 680°C. The other films were grown on $CaF_2$ and STO at 740°C.

**Laser-based polarimetry.** The ultrafast time-resolved polarimetry measurement is performed by using a Ti:Sa amplifier with 800 nm central wavelength, 40 fs pulse duration, and 1 kHz repetition rate. The laser is split into pump and probe beams. The pump beam, with 1.55 eV photon energy and vertical polarization, photoexcites the sample out of equilibrium. The subsequent probe beam, which is frequency-doubled to 400 nm (3.1 eV) with a Beta Barium Borate (BBO) crystal and horizontally-polarized, detects the transient photoinduced ellipticity change of the sample at a pump-probe delay time controlled by an optical delay stage. Both beams are at near normal incidence of the sample. The probe beam reflected from the sample is passed through a quarter-wave plate and a Wollaston prism, which spatially separates the s- and p-polarized probe beams. The difference of their intensities, $I_s$-$I_p$, is measured by a balanced photodetector, which is proportional to the change of the ellipticity angle. Such laser-based non-degenerate polarimetry technique is chosen here since it offers at least one order of magnitude higher sensitivity of the measured ellipticity signals in comparison to static or degenerate measurements[32]. The temperature-dependent ellipticity measurement is measured down to 4 K in a low temperature cryostat.

**Synchrotron x-ray techniques.** High-resolution x-ray diffraction and resonant x-ray scattering were carried out at beamline 6-ID-B of the Advanced Photon Source (APS) at Argonne National Laboratory. The x-ray probes the entire film depth, giving an average lattice parameter value. XRD shows total strain states at once and maximum position of the peak is selected to obtain lattice parameter of the tetragonal unit cell. Orthorhombicity is calculated from in-plane lattice parameters by fitting with the Pearson VII distribution function, which has been useful to handle peak shapes from conventional x-ray diffraction patterns. This combination of Lorentzian and Gaussian functions handles peak tails more appropriately than either Lorentzian or Gaussian, particularly in strained films. We consider the (228) tetragonal peak split into two orthorhombic peaks, which have the same full width at half maximum (FWHM) as the single tetragonal peak at room temperature. Regardless of any depth dependent strain gradient, the fit clearly determines peak splitting below the phase transition temperature. Atomic positions and bond angles were determined from the As-K edge anomalous x-ray scattering as described in the Supplementary Information.

**Data availability**

The data that support the findings of this study are available from the corresponding author upon reasonable request.

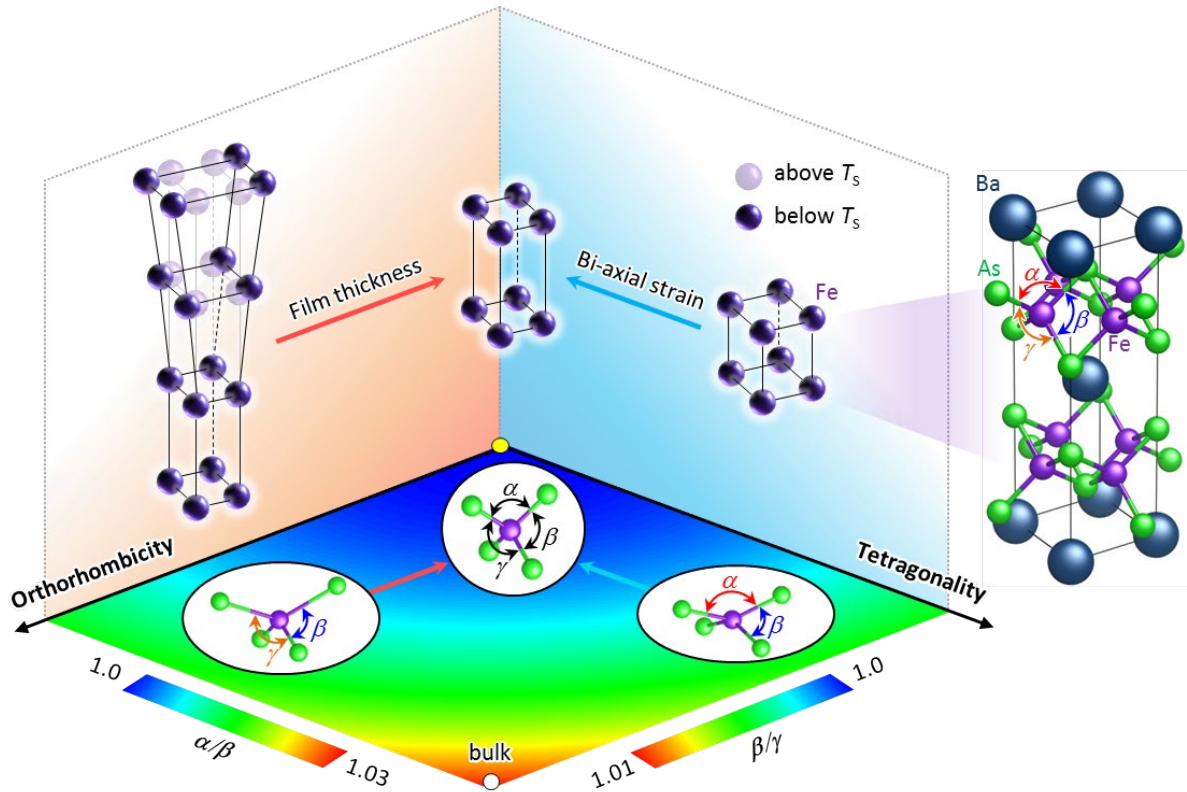

**Figure 1. Schematic picture of structural design in parent Fe-based compounds.** Individual control of orthorhombicity and tetragonality determines lattice stability toward or away from the perfect tetrahedral geometry in the presence of the tetragonal-to-orthorhombic transition. Each decoupled parameter is tuned by film thickness and bi-axial strain, independently. Orthorhombic transition below $T_s$ is more suppressed in thinner film by the substrate clamping effect and tetragonality is modified by in-plane strain. Orthorhombicity and tetragonality control anisotropy between angles $\beta$ and $\gamma$, and between $\alpha$ and $\beta$, respectively, and a perfect tetrahedron can be obtained by reducing structural anisotropy under control of orthorhombicity and tetragonality.

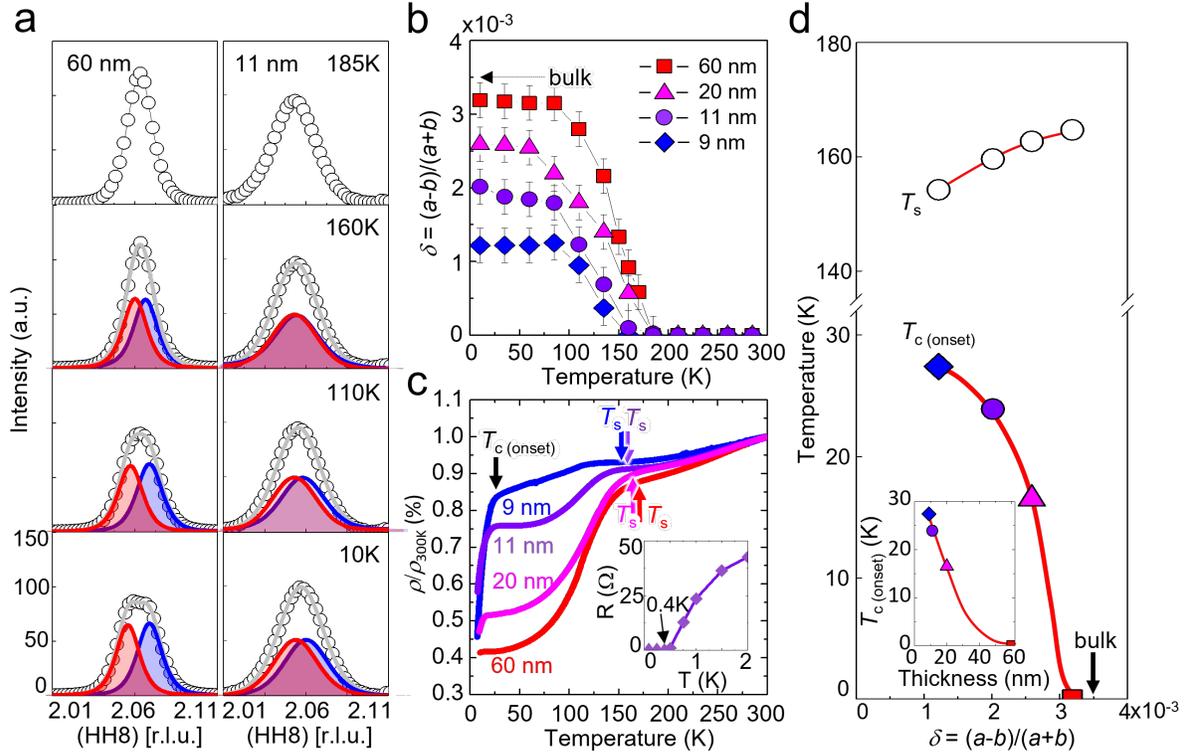

**Figure 2. Superconductivity enhanced by suppression of orthorhombic phases.** a) Tetragonal-to-orthorhombic structural change along the [hh8] direction across the tetragonal (228) reflection by x-ray diffraction at different film thickness. All films were grown on LiF substrates at 680℃. b) Temperature-dependent orthorhombicity $\delta$ extracted from (228) reflections. c) Normalized resistivity as a function of temperature for different film thicknesses. Inset (11 nm, LiF substrate) shows full superconducting transition with zero resistance at 0.4K. d) Superconducting $T_c$ and structural phase transition ($T_s$) correlated with $\delta$ (taken at 10K). Note that $c/a$ is fixed as 3.348 below $T_s$. $T_c$ as a function of film thickness shown in the inset. Red lines are guides to the eye.

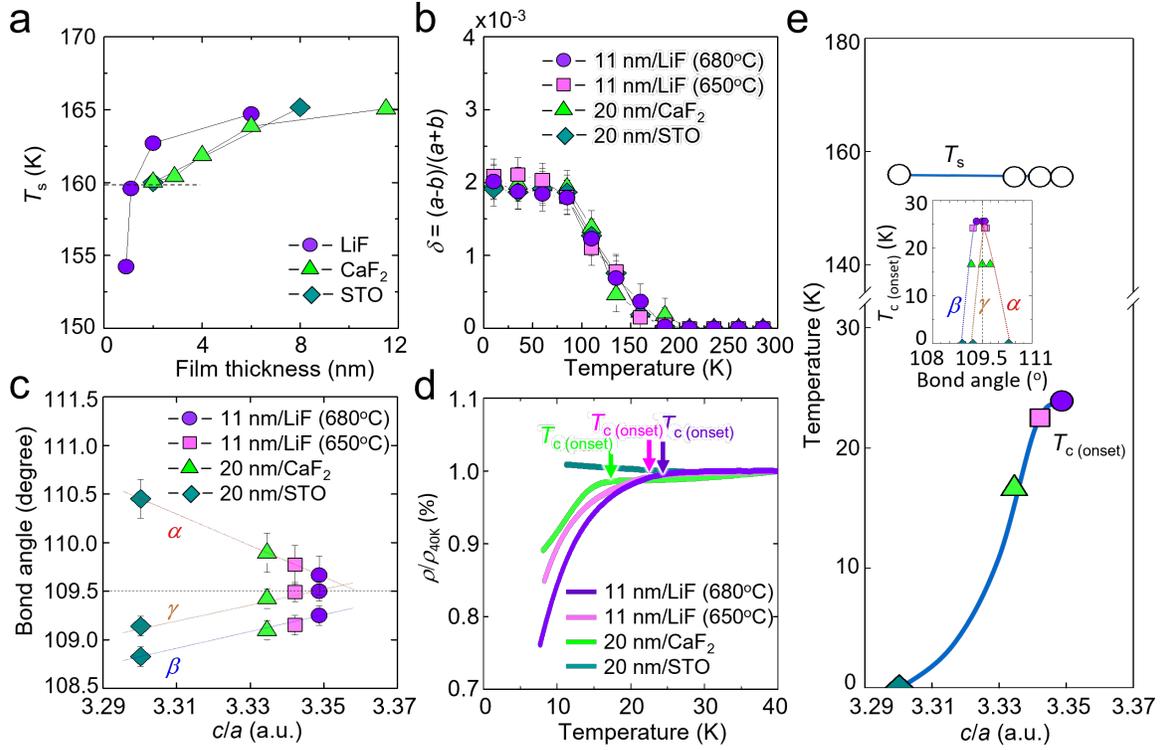

**Figure 3. Superconductivity tuned by tetragonal structures.** a) Thickness-dependent $T_s$ in different substrates. Growth temperature on LiF is 680°C. b) Identical temperature-dependent $\delta$ of films with the same $T_s$. c) As-Fe-As bond angles $\alpha$, $\beta$ and $\gamma$ in tetrahedron controlled by $c/a$ ratio. d) Normalized resistivity showing different onset of $T_c$. e) $T_c$ and $T_s$ as a function of tetragonality ($c/a$ taken at 10K). $T_c$ as a function of bond angles shown in the inset. Blue lines are guides to the eye.

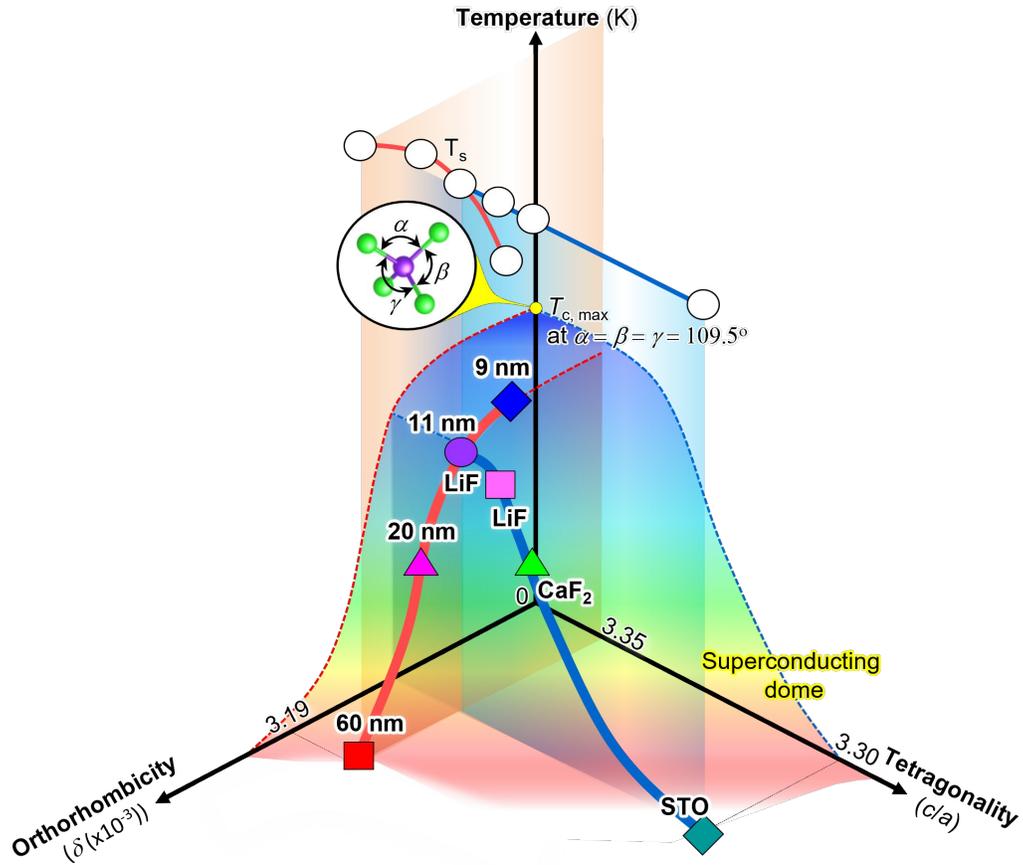

**Figure 4. Three-dimensional phase diagram of parent BaFe$_2$As$_2$ materials.** Demonstration of explicit guidelines showing how to enhance $T_c$ with the existence of tetragonal-to-orthorhombic transition. Decoupling the two structural parameters of orthorhombicity and tetragonality is the key for approaching the ideal tetrahedron where the three bond angles are identical (yellow dot). Solid red and blue lines are drawn from the measured superconducting $T_c$ under independent control of orthorhombicity and tetragonality.

AUTHOR INFORMATION

**Corresponding Author**

*eom@engr.wisc.edu (C. B. Eom)

**Author contributions**

J.H.K. and C.B.E. conceived the project. C.B.E., M.S.R., R.M., J.W., and E.H. supervised the experiments. J.H.K. synthesized the samples and carried out the x-ray diffraction and transport measurements. J.H.K., P.J.R. and J.W.K. performed synchrotron x-ray diffraction and resonant scattering. J.S., J.P. and T.H.K. helped with the synchrotron experiments. N.C., J.S., M.S.R. and R.M. carried out transport measurements. L.L., D.C. and J.W. performed laser-based polarimetry to measure ultrafast ellipticity signals. Y.G.C. and E.E.H prepared PLD targets for the thin film growth. J.H.K. and C.B.E. wrote the manuscript. C.B.E. directed the research.

**Acknowledgements**

This work was supported by the US Department of Energy (DOE), Office of Science, Office of Basic Energy Sciences (BES), under award number DE-FG02-06ER46327 (C.B.E.). This research used resources of the Advanced Photon Source, a U.S. Department of Energy (DOE) Office of Science User Facility operated for the DOE Office of Science by Argonne National Laboratory under Contract No. DE-AC02-06CH11357. The synchrotron x-ray diffraction work was partially supported by NSF through the University of Wisconsin Materials Research Science and Engineering Center (DMR-1720415). The work at NHMFL was supported under NSF Cooperative Agreement DMR-1644779 and by the State of Florida. The optical polarimetry work (L.L., D.C. and J.W.) was supported by the U.S. Department of Energy, Office of Basic Energy Science, Division of Materials Sciences and Engineering (Ames Laboratory is operated for the U.S. Department of Energy by Iowa State University under Contract No. DE-AC02-07CH11358). We thank Ian Fisher for helpful discussion.

**Competing Interests**

The authors declare no competing interests.

**Additional information**

Supplementary information is available in the online version of the paper. Reprints and permissions information is available online at www.nature.com/reprints. Correspondence and requests for materials should be addressed to C.B.E.

# Supplementary Information

## Local atomic configuration control of superconductivity in the undoped pnictide parent compound $BaFe_2As_2$

Jong-Hoon Kang[1], Philip J. Ryan[2,5], Jong-Woo Kim[2], Jonathon Schad[1], Jacob P. Podkaminer[1], Neil Campbell[3], Joseph Suttle[3], Tae Heon Kim[1], Liang Luo[4], Di Cheng[4], Yesusa G. Collantes[6], Eric E. Hellstrom[6], Jigang Wang[4], Robert McDermott[3], Mark S. Rzchowski[3], Chang-Beom Eom[1]*

[1]Department of Materials Science and Engineering, University of Wisconsin-Madison, Madison, Wisconsin 53706, USA

[2]Advanced Photon Source, Argonne National Laboratory, Argonne, Illinois 60439, USA

[3]Department of Physics, University of Wisconsin-Madison, Madison, Wisconsin 53706, USA

[4]Department of Physics and Astronomy, Ames Laboratory, Iowa State University, Ames, Iowa 50011, USA

[5]School of Physical Sciences, Dublin City University, Dublin 9, Ireland

[6]Applied Superconductivity Center, National High Magnetic Field Laboratory, Florida State University, 2031 East Paul Dirac Drive, Tallahassee, FL 32310, USA

*eom@engr.wisc.edu

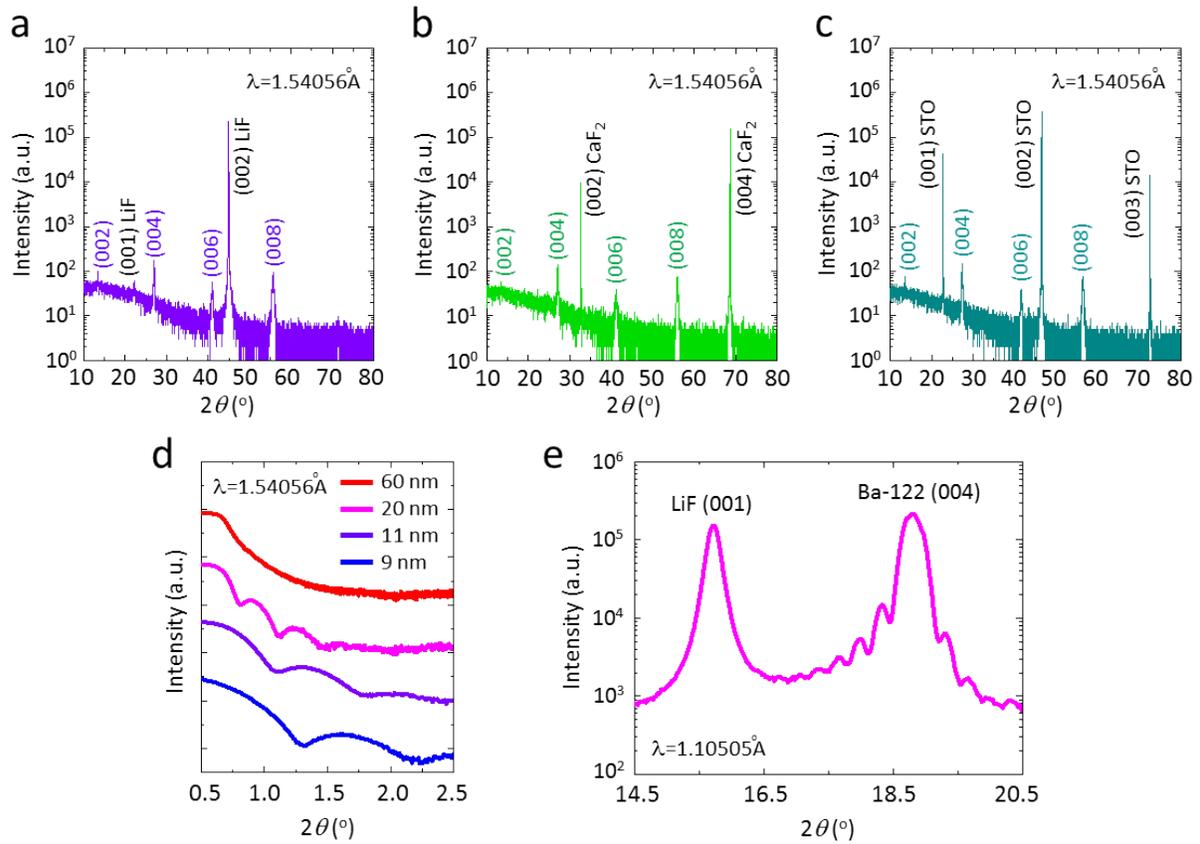

**Figure S1. Epitaxial quality of Ba-122 films.** Out-of-plane $\theta$–$2\theta$ XRD patterns of the 20 nm thick films on a) LiF (680°C), b) CaF$_2$ (740°C), and c) STO (740°C) substrates. d) X-ray reflectivity spectra of different film thicknesses with LiF substrates. e) X-ray diffraction patterns showing fringe spacing around Ba-122 (004) reflection on LiF (20 nm, 680°C).

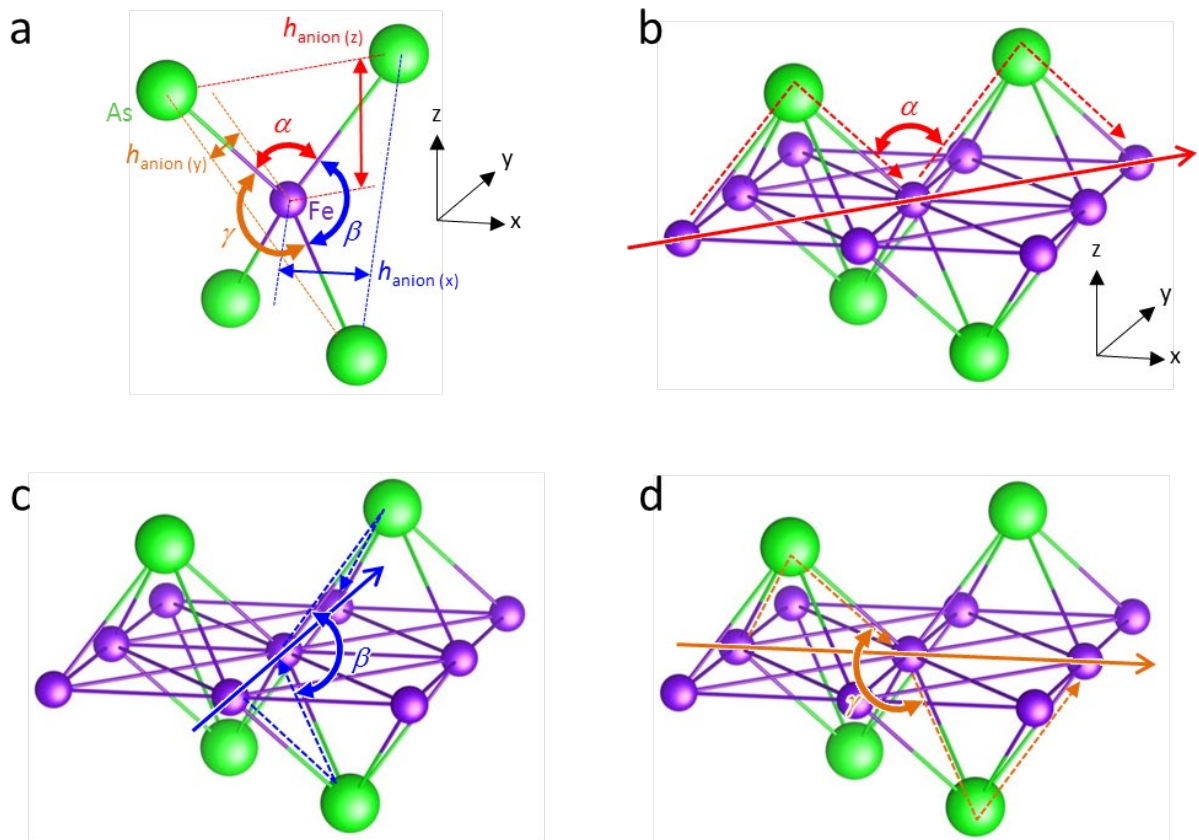

**Figure S2. As-Fe-As tetrahedron geometry.** a) Bond angles determined by anion heights in the tetrahedron. b) Next-nearest neighbor interaction via bond angle $\alpha$ along the diagonal direction in the orthorhombic notation. Nearest neighbor hoppings via c) bond angle $\beta$ along the $y$ direction and d) bond angle $\gamma$ along the $x$ direction.

**Extraction of $T_s$ and $T_c$ from resistivity in broad phase transition.** $T_s$ and $T_n$ are extracted from the first derivative of the resistivity to define the structural transition ($T_s$) and antiferromagnetic transition temperatures ($T_n$). Even though $T_s$ and $T_n$ of thin film are not as clear as bulk values, we can determine them from the onset and peak position of dρ/dT, respectively (Figure S3).

The only caveat that should be mentioned is that the variation of the effective elastic moduli of the thin film (in particular of the effective $c_{66}$) as a function of vertical distance from the interface perforce results in a broadening of signatures associated with $T_s$ observed in both x-ray diffraction and also in transport measurements for thicker films. This broadening occurs because the structural phase transition occurs at a temperature that is renormalized from the bare nematic critical temperature due to coupling to the crystal lattice; the degree to which $T_s$ is renormalized depends on the stiffness of the lattice[1-3]. Equally, any strain-dependence of the superconducting critical temperature $T_c$ will also result in a broadened transition due to the strain gradient, with the onset of zero resistivity determined by the first part of the material to form a filamentary superconducting pathway (Figure S5). However, as we will show, even the average values of $\delta$ and of $T_s$ for such films which are probed by x-ray diffraction show a clear systematic trend as a function of film thickness, permitting correlation between the average values of bond angles $\alpha$ and $\beta$ for a given film and the onset of $T_c$ measured at low temperatures.

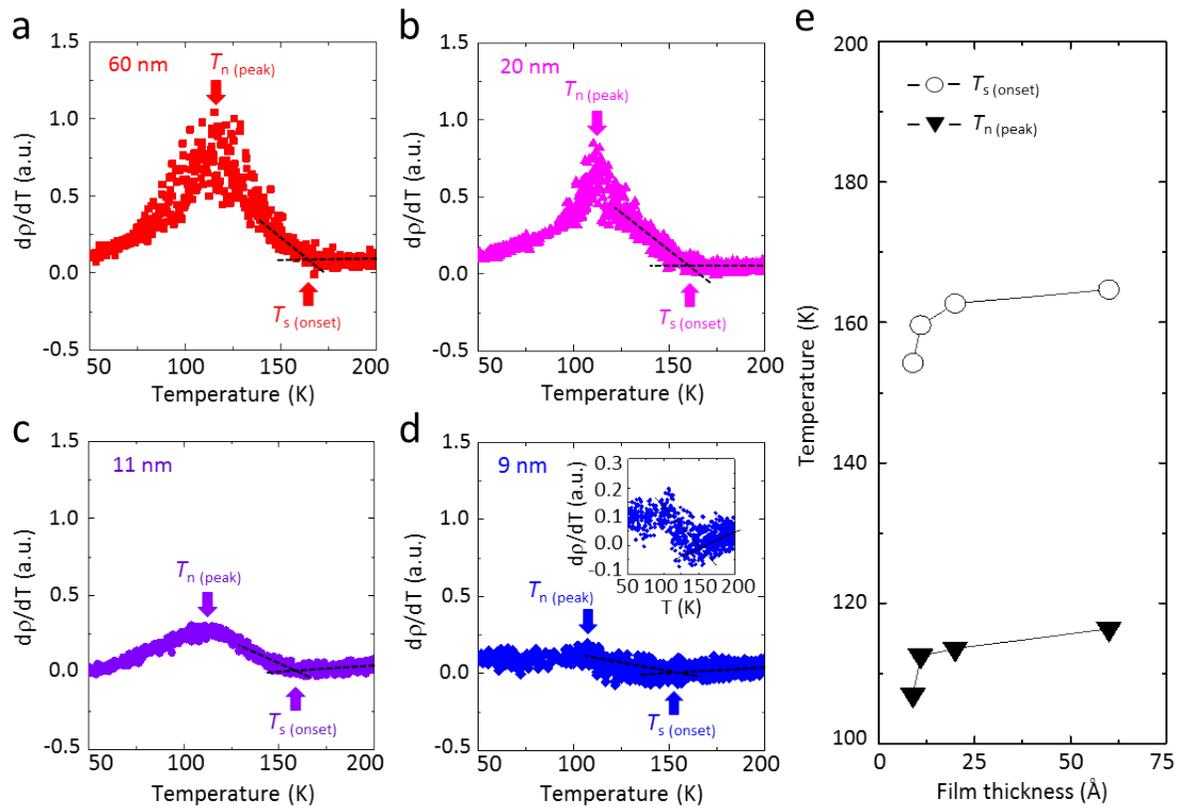

**Figure S3. $T_s$ and $T_n$ controlled by film thickness.** $T_s$ and $T_n$ extracted from dρ/dT in a) 60 nm, b) 20 nm, c) 11 nm, and d) 9 nm thick films grown on LiF substrates at 680°C. e) Thickness-dependent $T_s$ and $T_n$.

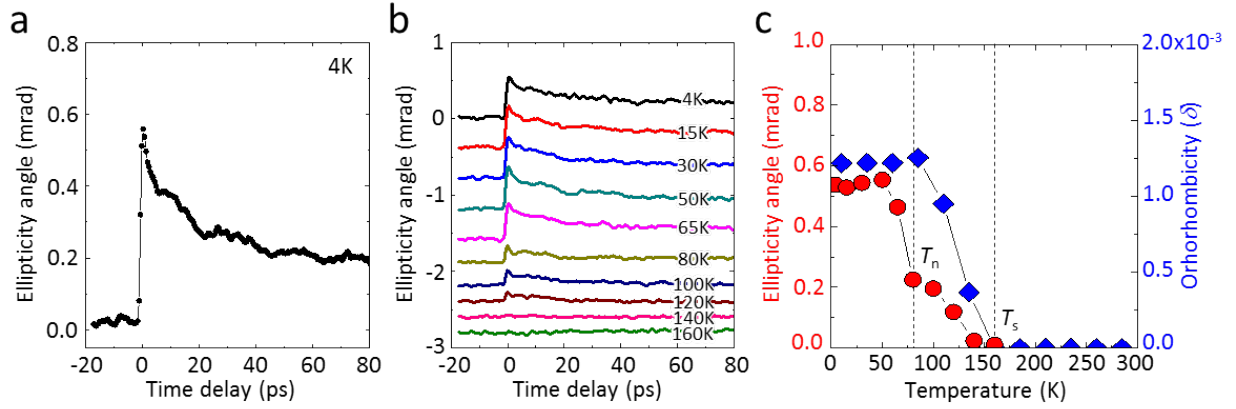

**Figure S4. Anisotropic normal-state characteristics probed by ultrafast ellipticity and structural orthorhombicity methods.** a) The representative photoinduced ellipticity signal of the 9 nm sample (LiF substrate) at $T$=4 K. The 800 nm (1.55 eV) pump beam has 2 mJ/cm$^2$ fluence and vertical polarization. The 400 nm (3.1 eV) probe beam has horizontal incident polarization. The signal is directly proportional to the two-fold in-plane anisotropy in BaFe$_2$As$_2$ samples. b) Detailed temperature dependence of the photoinduced ellipticity dynamics of the 9 nm sample with the same experimental conditions as in a). Note the traces are offset for clarity. c) Photoinduced ellipticity amplitude and orthorhombicity as a function of temperature. $T_s$ values are the same as measured by both laser-based polarimetry technique and synchrotron X-ray technique, which indicates the emergence of orthorhombicity. In addition, the measured ellipticity signals exhibit a kink at $T_n$ due to the magnetic feedback effect[4].

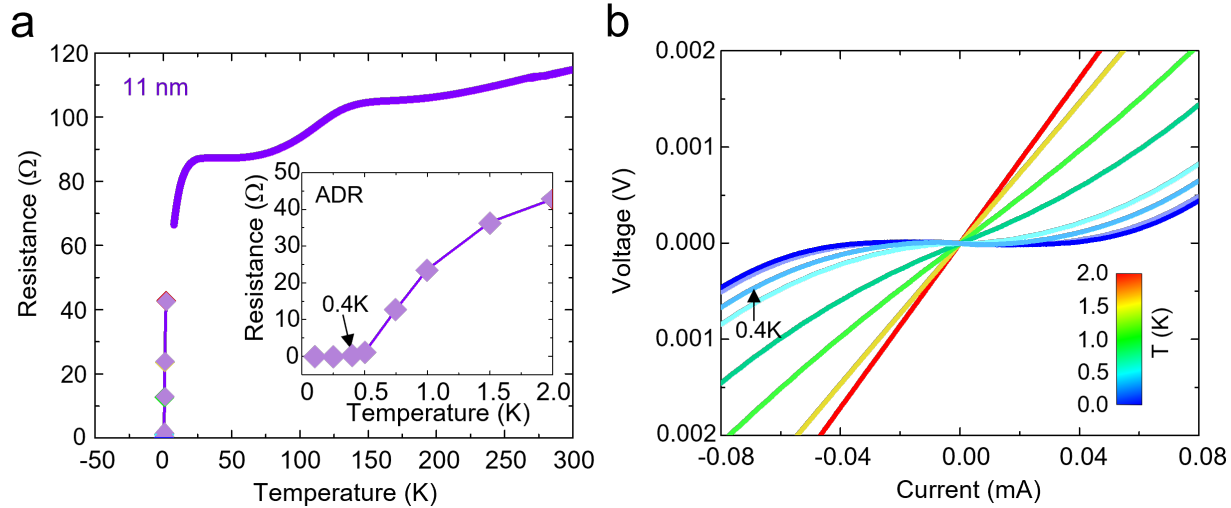

**Figure S5. Superconductivity with zero resistance.** a) Temperature-dependent resistance of 11 nm thick film on LiF substrate (680 °C). Inset shows $T_c$ = 0.4 K measured in adiabatic demagnetization refrigerator. b) Current-voltage characteristics indicating zero resistance at 0.4K.

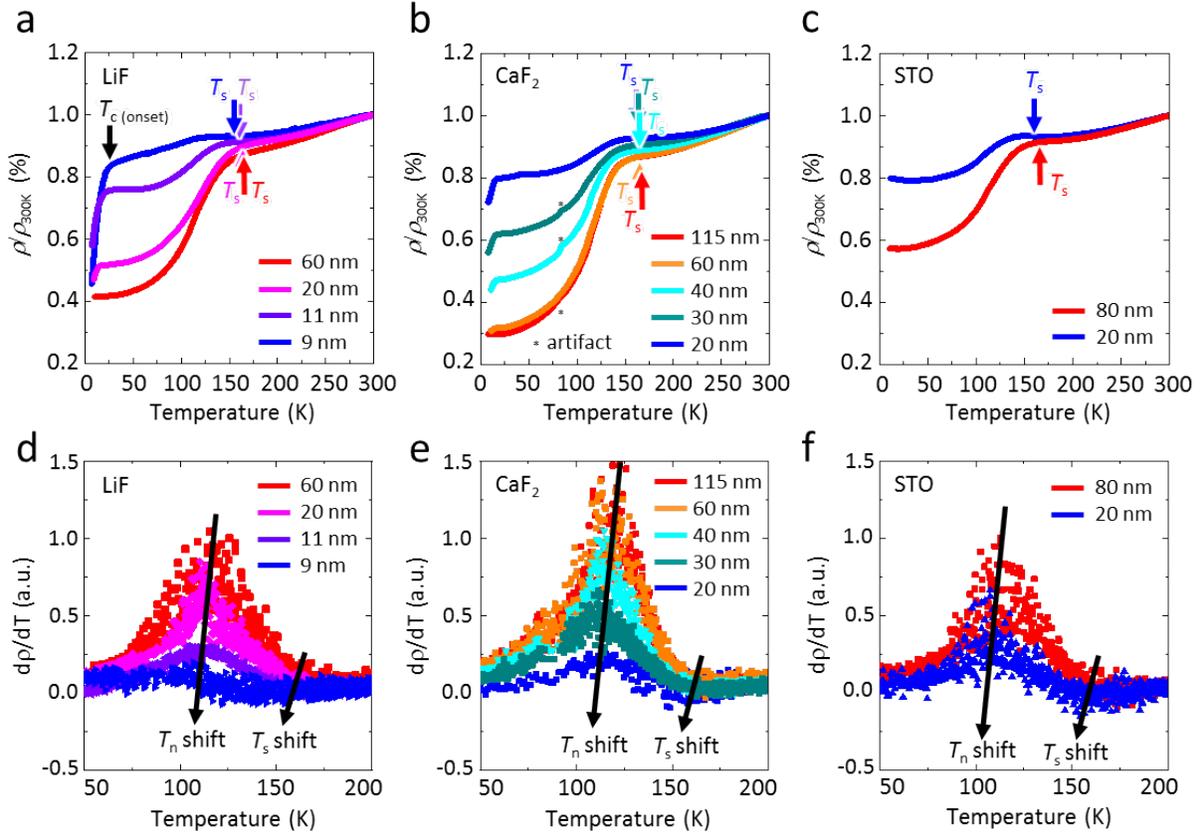

**Figure S6. Temperature-dependent normalized resistivity and the first derivative of the resistivity of the films.** $T_s$ shifts to lower temperature as film thickness decreases in a) LiF (680°C), b) CaF$_2$ (740°C), and c) STO (740°C) substrates. The first derivative of the resistivity of Ba-122 films on d) LiF, e) CaF$_2$, and f) STO substrates.

**Tetrahedron design in thin films to overcome bulk limitations.** Previous phase diagram was a limited observation by external variables and unfortunately the highest $T_c$ shown in the superconducting dome is not always the maximum value we can reach in a given material because of physical limitations of bulk study[5]. First, the external variables (e.g., hydrostatic pressure, doping) cannot separate orthorhombicity and tetragonality[6]. Second, the isotropic pressure is not adequate to control tetragonality ($c/a$), and sometimes tetragonal structure can be collapsed under pressure[7,8]. Third, pressure cannot expand materials, but only compress them, which means we cannot always achieve the ideal structure[9]. However, our approach in thin film overcomes these obstacles to obtain systematic change of lattice parameters, even in the presence of a tetragonal-to-orthorhombic transition (Figure S7).

The internal "height" of the As ion above the Fe plane in the unit cell (parameterized by the dimensionless quantity $z$) is necessary to calculate the anion height for the bond angle $\alpha$ in the tetrahedron. The other bond angles, $\beta$ and $\gamma$, can be calculated by lattice parameters (see Methods). The internal parameter $z$ is not directly controlled by lattice parameters, but is found to be determined by Coulomb interaction with different nearest neighbor atoms[10,11]. The internal parameter defining the "height" of the As atom in the unit cell, $z$, were obtained from a combination of high resolution x-ray diffraction and resonant scattering (Figure S8).

Atomic positions and bond angles were determined from the As-K edge anomalous x-ray scattering as described below. The symmetry group of Ba-122 is I4/mmm. Both Ba and Fe are at the special Wyckoff positions 2a and 4d respectively I4/mmm crystal symmetry. If the crystal maintains this symmetry with the application of biaxial strain, the relative positions of Ba and Fe are fixed within the given unit cell. On the other hand, the As position is 4e so that the relative $z$ position (along the $c$ axis) in the unit cell can vary and not break symmetry. For the bulk crystal case under hydrostatic pressure, only As's relative $z$ position changes[41-43]. Therefore, any intensity difference measured in the (00L) Bragg reflections is only sensitive to changes in As's relative $z$ position. Since the proportion of As contribution to these reflections is significant, one can expect to determine the absolute position of the As atom from the As-$K$ edge anomalous x-ray scattering.

The scattering amplitude of the atom can be written in the form

$$f(Q,\omega) = f^0(Q) + f'(\omega) + if''(\omega)$$

where $f'$ and $f''$ are the real and imaginary parts of the dispersion corrections, which are energy dependent at absorption edges as the resonant scattering terms. Using imaginary $f''$ obtained from fluorescence data, real $f_{As}$ was calculated from the (006) structure factor from the measured intensity of the (006) reflection. Here, the structure factor was used for the given As relative position so that $f'$ can vary when you change the relative height. However, it does not change much because (006) is not sensitive to the As $z$ position. Therefore, we could calculate $f_{As}$ as a constant and the calculated $f_{As}$ was generated to match the energies for $f''$, with 3 eV offset for convenience.

$$f_{As}(Q,\omega) = f^0(Q) + f'(\omega) + if''(\omega)$$
$$= \left(F_{(006)} - f_{Ba} \cdot \sum e^{-2\pi i(6\times(0+0.5))} - f_{Fe} \cdot \sum e^{-2\pi i(6\times(0.25+0.75))}\right) / \sum e^{-2\pi i(6z)}$$

With the $f_{As}$, other (008) reflections were calculated and compared with the measured (008)

reflection of films on various substrates (Figure S7).

$$F_{(008)} = f_{As}(Q,\omega) \cdot \sum e^{-2\pi i(8z)} + f_{Ba} \cdot \sum e^{-2\pi i(8\times(0+0.5))} + f_{Fe} \cdot \sum e^{-2\pi i(8\times(0.25+0.75))}$$

For all the films with different strain states, the relative As contribution to the (00L) reflections does not change. From the (008) energy scans shown in Supplementary Figure S8, the relative As $z$ positions in these samples are 0.3552. To extract estimated $z$ position changes, a series of calculated energy scans is plotted which can be employed as an As displacement scale, from which a tabulated angle can be estimated to help refine the angles extracted from the tetragonal unit cell changes.

With the combination of x-ray diffraction and resonant scattering, lattice information and the As relative $z$ position were obtained and the anion height and As-Fe-As bond angle can be calculated (Figure S8).

$$h_{anion(\alpha)} = c \times (z - 0.25)$$

$$h_{anion(\beta)} = 0.25 \times a$$

$$h_{anion(\gamma)} = 0.25 \times b$$

$$\alpha = 2 \times \tan^{-1}\left(\frac{\sqrt{\left(\frac{a}{4}\right)^2 + \left(\frac{b}{4}\right)^2}}{h_{anion(\alpha)}}\right) = 2 \times \tan^{-1}\left(\frac{\sqrt{\left(\frac{a}{4}\right)^2 + \left(\frac{b}{4}\right)^2}}{c \times (z - 0.25)}\right)$$

$$\beta = 2 \times \tan^{-1}\left(\frac{\sqrt{\left(\frac{b}{4}\right)^2 + \left(cz - \frac{c}{4}\right)^2}}{h_{anion(\beta)}}\right) = 2 \times \tan^{-1}\left(\frac{\sqrt{\left(\frac{b}{4}\right)^2 + \left(cz - \frac{c}{4}\right)^2}}{0.25 \times a}\right)$$

$$\gamma = 2 \times \tan^{-1}\left(\frac{\sqrt{\left(\frac{a}{4}\right)^2 + \left(cz - \frac{c}{4}\right)^2}}{h_{anion(\gamma)}}\right) = 2 \times \tan^{-1}\left(\frac{\sqrt{\left(\frac{a}{4}\right)^2 + \left(cz - \frac{c}{4}\right)^2}}{0.25 \times b}\right)$$

where $a$, $b$, $c$ are the orthorhombic lattice parameters and $z$ is the As relative position.

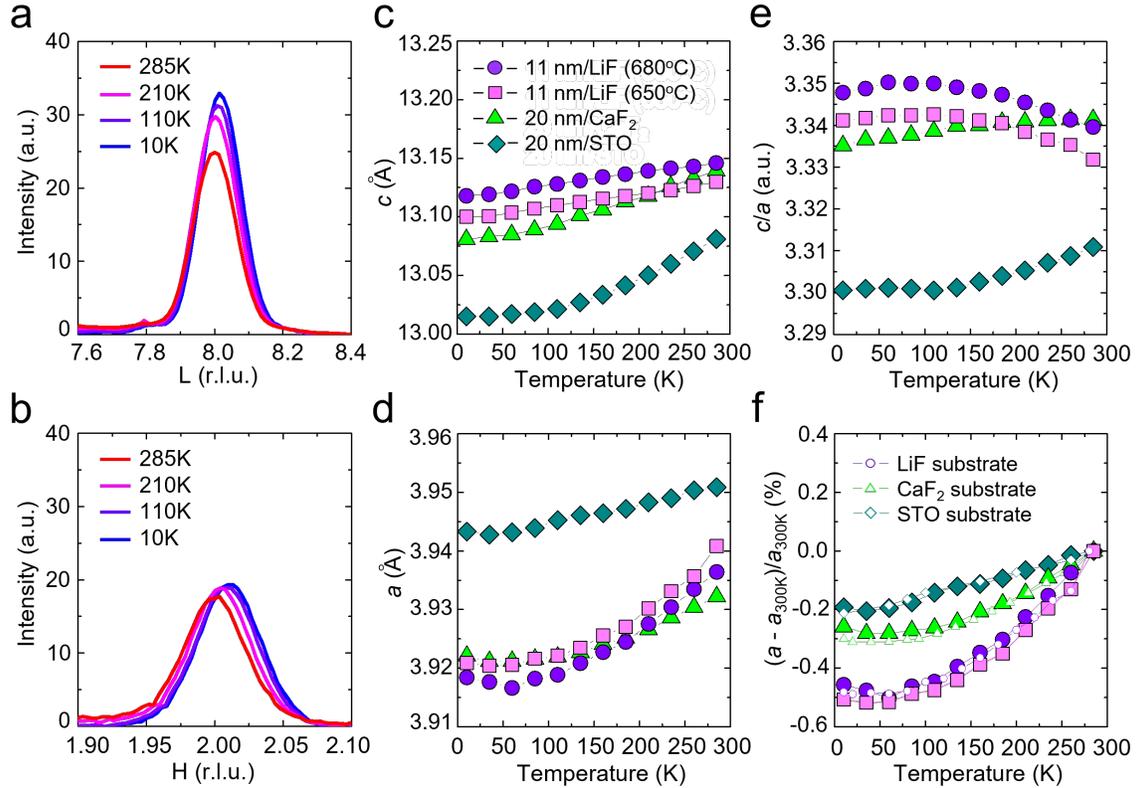

**Figure S7. Lattice parameters and tetragonality from x-ray diffraction.** a) Temperature-dependent out-of-plane (008) and b) in-plane (228) reflections of the Ba-122 films (STO substrates, 740°C). c) Out-of-plane and d) in-plane lattice parameters of the films on various substrates in tetragonal notation as a function of temperature. e) Calculation of tetragonality from $c$ and $a$ lattice parameters. f) Percentage of in-plane lattice change of the films, compared with the percentage change of the substrates. The changing rate of the film lattice is driven by thermal contraction of the substrates.

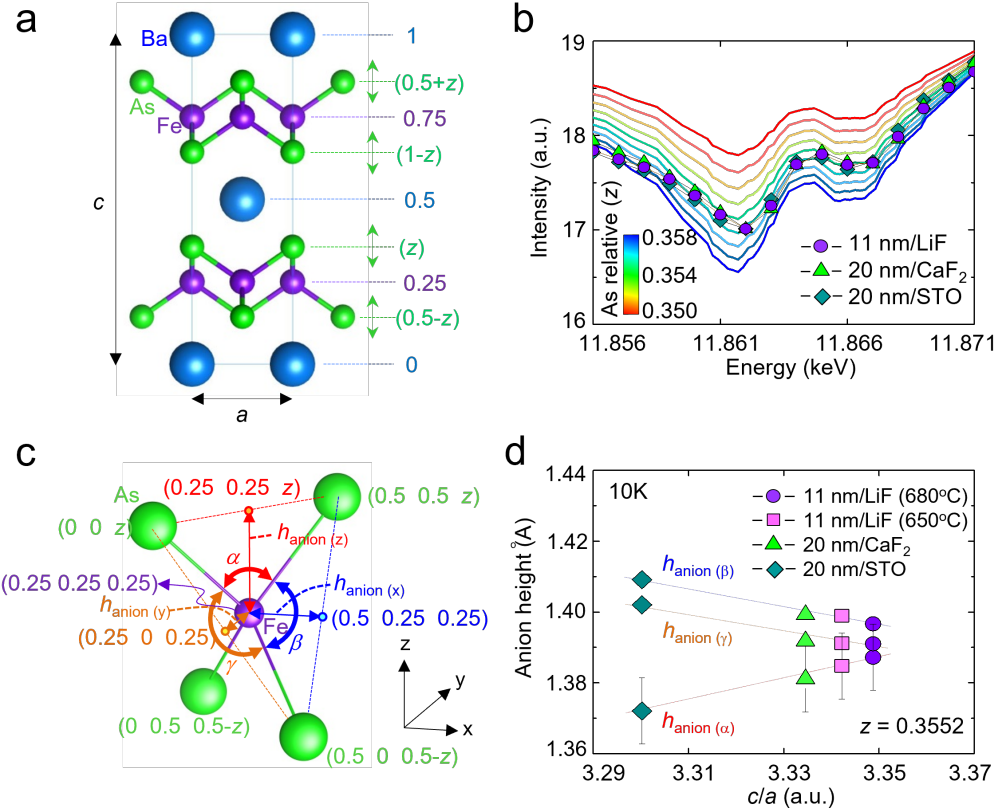

**Figure S8. Crystal structure and tetrahedron geometry for the calculation of As anion height.**
a) Ba-122 tetragonal unit cell and relative $z$ position of Ba, Fe, and As atoms. b) Energy scan for resonant scattering to determine As relative $z$ position by comparing calculated and measured intensity spectra. c) Relative positions of Fe and As atoms and three anion heights in tetrahedron with orthorhombic notation. d) Anion height controlled by tetragonality.